# Negligible electronic contribution to heat transfer across intrinsic metal/graphene interfaces

Bin Huang[1], Yee Kan Koh[1]

[1]*Department of Mechanical Engineering, and Centre for Advanced 2D Materials, National University of Singapore, Singapore*

**ABSTRACT**

Despite the importance of high thermal conductance (i.e., low thermal resistance) of metal contacts to thermal management of graphene devices, prior reported thermal conductance ($G$) of metal/graphene interfaces are all relatively low, only 20-40 MW m$^{-2}$ K$^{-1}$. One possible route to improve the thermal conductance of metal/graphene interfaces is through additional heat conduction by electrons, since graphene can be easily doped by metals. In this paper, we evaluate the electronic heat conduction across metal/graphene interfaces by measuring the thermal conductance of Pd/transferred graphene (trG)/Pd interfaces, prepared by either thermal evaporation or radio-frequency (rf) magnetron sputtering, over a wide temperature range of $80 \leq T \leq 500$ K. We find that for the samples prepared by thermal evaporation, the thermal conductance of Pd/trG/Pd is 42 MW m$^{-2}$ K$^{-1}$. The thermal conductance only weakly depends on temperature, which suggests that heat is predominantly carried by phonons across the intrinsic Pd/graphene interface. However, for Pd/trG/Pd samples with the top Pd films deposited by rf magnetron sputtering, we observe a significant increment of thermal conductance from the intrinsic value of 42 MW m$^{-2}$ K$^{-1}$ to 300 MW m$^{-2}$ K$^{-1}$, and $G$ is roughly proportional to $T$. We attribute the enhancement of thermal conductance to an additional channel of heat transport by electrons via atomic-scale pinholes formed in the graphene during the sputtering process. We thus conclude that electrons play a negligible role in heat conduction across intrinsic interfaces of metal and pristine graphene, and the contribution of electrons is only substantial if graphene is damaged.

**TEXT**

Graphene, with its exceptional electrical and thermal properties,[1-3] is actively investigated for a wide range of emerging applications, including for example in electronics,[4] optoelectronics,[5] sensors,[6] and flexible electronics.[7] One of the main challenges for practical applications of graphene devices is the high electrical and thermal contact resistance between graphene and metal electrodes.[8-10] For evaporated metal contacts on graphene, the area specific contact resistivity ($\rho_c$) is on the order of $10^{-6}$ of $\Omega$ cm$^{-2}$,[11, 12] which is orders of magnitudes higher than the resistivity ($\approx 10^{-9}$ $\Omega$ cm$^{-2}$) of metal contacts in conventional Si-based devices.[13] Over the past years, substantial efforts have been made to reduce the electrical contact resistivity of metal/graphene contacts, through e.g., light plasma treatment,[14] ultraviolet/ozone (UVO) treatment,[15, 16] nickel-etched treatment,[17] sputtering,[18] and contact area patterning.[19] However, less progress was made in the improvement of the thermal properties[20] of metal contacts. Despite prior efforts, the thermal conductance ($G$) of metal contacts on graphene is still fairly low (20-40 MW m$^{-2}$ K$^{-1}$) compared to other epitaxial solid/solid interfaces ($\approx$700 MW m$^{-2}$ K$^{-1}$),[21] and as a result, heat dissipation from small (<500 nm) graphene devices[22] is seriously impeded.

In principle, an additional heat transport channel could be facilitated by electronic transport across graphene/metal interfaces, since metals are known to induce a high concentration of electrons in graphene.[23] Former study by our group on heat transfer across aluminum/transferred graphene/copper (Al/trG/Cu) interfaces, however, indicate no substantial electronic heat transport even when graphene is as-grown on copper foils by chemical vapor deposition (CVD), and a significant concentration of charge carriers ($\approx 3\times 10^{12}$ cm$^{-2}$) is induced in graphene by metals.[24] Despite the null results, contribution of electrons cannot be ruled out in intrinsic metal/graphene interfaces, because copper substrates in the previous study were exposed to atmospheric conditions during sample preparation and thus a thin layer of native oxide ($\approx 0.5$ nm) was inevitably formed. This thin layer of metal native oxide could be sufficient to inhibit transmission of electrons across the metal/graphene interfaces.

In this paper, we select palladium (Pd) which does not oxidize under atmospheric conditions, to study the electronic heat transport across intrinsic metal/graphene interfaces. We find that for intrinsic Pd/graphene/Pd interfaces prepared by thermal evaporation, electrons do not play an active role in the heat conduction across metal contacts on pristine graphene. Together with our previous finding that scattering of electrons by remote interfacial phonons (RIP) does not significantly enhance thermal conductance of graphene/SiO$_2$ interfaces,[25] we conclude that additional heat transport by electrons is not a viable route to enhance the thermal conductance of pristine graphene interfaces. Interestingly, we find that the thermal conductance $G$ of the sandwiched structure prepared by radio-frequency (rf) magnetron sputtering is enhanced from 42 MW m$^{-2}$ K$^{-1}$ to 300 MW m$^{-2}$ K$^{-1}$. We attribute the enhancement to electronic heat transport via atomic-scale pinholes generated in the graphene during the sputtering treatment.

Our samples consist of transferred graphene (trG) sandwiched between two layers of deposited Pd films on GaN/sapphire, see Figure 1a. We choose GaN/sapphire as the substrate for this study because the accuracy of our thermal measurements of Pd/trG /Pd interfaces is substantially improved with the high thermal conductivity of the GaN substrates (independently measured as 180 W m$^{-1}$ k$^{-1}$). The top Pd film (≈100 nm) of one sample was deposited by thermal evaporation, while the top Pd films of two samples were deposited by rf magnetron sputtering at a deposition rate of 0.3 Å/s or 1 Å/s, with an argon pressure of 3 mTorr. The bottom Pd films, deposited by rf magnetron sputtering, are rather thick (300-400 nm). The thick bottom Pd films reduce the sensitivity of the thermal measurements to the thermal conductance of the bottom Pd/GaN interfaces, see more discussion below.

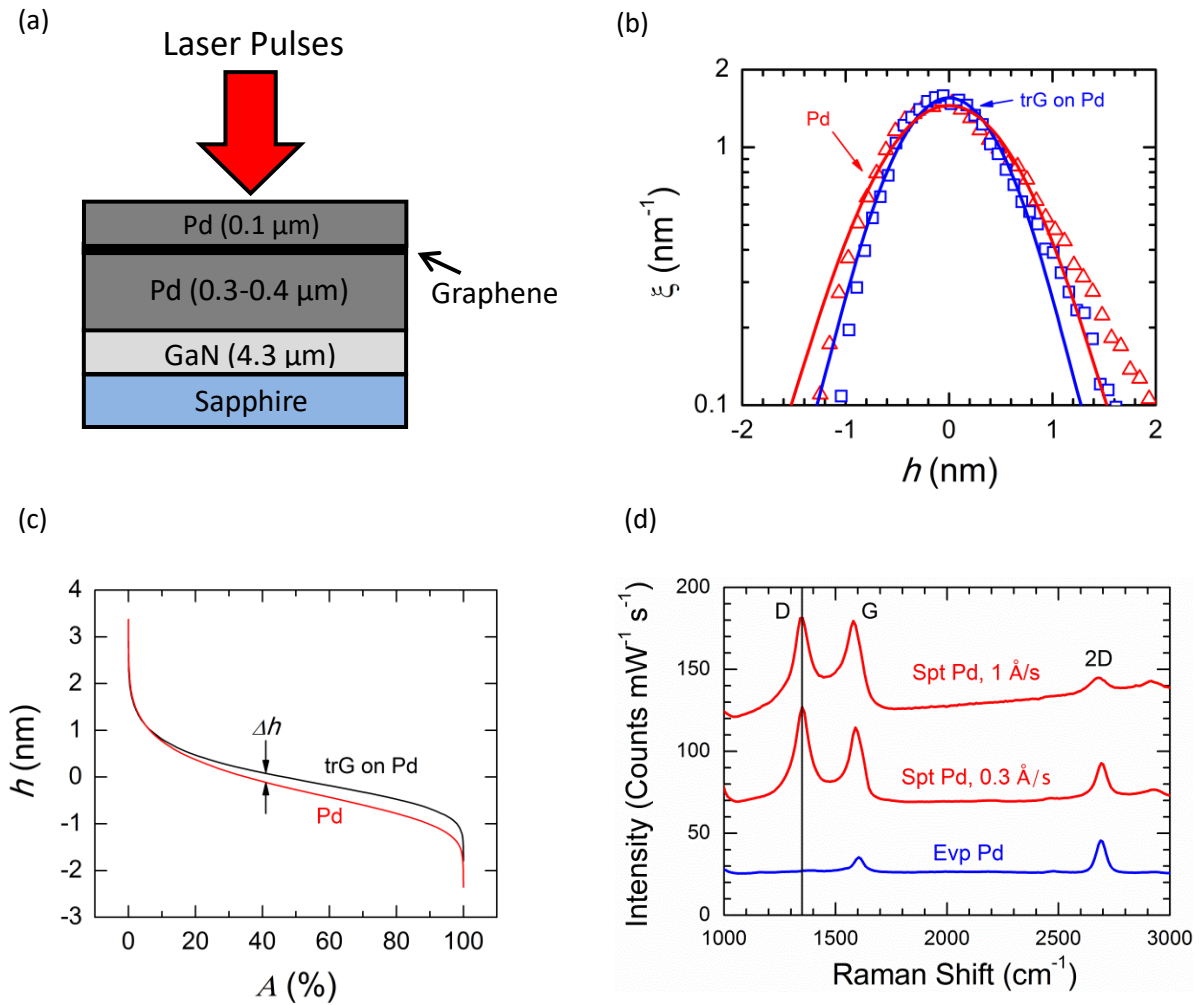

**Figure 1.** (a) Cross-section schematic diagram of our Pd/graphene/Pd samples. (b) Depth histogram of transferred graphene on an evaporated Pd thin film (trG/Pd, open square) and the evaporated Pd thin film alone (Pd, open triangle) derived from AFM topographic images. A Gaussian function is used to fit the depth histograms. ξ is a spatial frequency that we defined in reference [25]; by plotting ξ, we have $\int_{-\infty}^{\infty} \xi dh = 1$ and thus the depth histogram does not depend on the height interval assumed to derive the plot. (c) Relative height of the transferred graphene on Pd and the Pd substrate derived from AFM depth histogram in Fig. 1b, plotted as a function of accumulative percentage of area *A*, see the main text for the definition of *A*. We define the graphene to be conformal when $\Delta h < 0.5$ nm. (d) Raman spectra of CVD graphene transferred on $SiO_2$ coated with approximately 5 to 10 nm thick Pd film, deposited either by thermal evaporation (blue) or by rf magnetron sputtering (red) using a power density of 1.32 and 11.4 W $cm^{-2}$ at an Argon pressure of 3 mTorr, with a deposition rate of 0.3 Å/s and 1 Å/s, respectively, as labeled. The spectra are vertically shifted by multiples of 50 counts $mW^{-1}$ $s^{-1}$ for clarity.

Our graphene, grown on copper foils by chemical vapor deposition (CVD), were purchased from Graphene Supermarket. During the preparation of our samples, we follow procedures stated in reference 25 to transfer the graphene unto the bottom Pd films. We spin-coat poly(bisphenol A carbonate) (PC) on graphene as the support layers, because PC is easier to completely dissolve in chloroform. We etch away the copper substrate by floating the graphene samples on a 7 wt. % ammonium persulfate (APS) solution. After the graphene is cleaned and transferred to the thick bottom Pd films, we bake dry, and immerse the samples in chloroform for 24 hours to strip away the PC layer. Finally, we deposit the ≈100 nm thick Pd films on the graphene samples by either rf magnetron sputtering or thermal evaporation.

We employ tapping mode atomic force microscopy (AFM) to characterize the cleanliness and conformity of the graphene after the transfer process. For our samples, we are not able to verify whether the transfer is clean from the AFM topography images, because the underlying Pd films are fairly rough with a root-mean-square (rms) roughness of 2.1 nm, and thus PC residues are not easily differentiable from the intrinsic roughness, see Figure S1a. Instead, we use AFM phase images to identify PC residues,[24] and observe no significant phase difference and thus no significant amount of PC residues, see Figure S1b. We also quantify conformity of graphene to the substrates from the AFM depth images, using an approach similar to that described in reference 25. To quantify conformity of graphene, we derive the relative height $h$ of our samples before and after graphene transfer from the depth histogram in Fig. 1b and plot it as a function of the accumulative percentage of area, see Figure 1c. The accumulative percentage of area $A(h)$ is defined as the total areas of the AFM topographic images with a relative height higher than $h$. We estimate difference in the relative height $\Delta h$ of graphene and substrates, and consider graphene to be conformal if $\Delta h < 0.5$ nm. For all our samples, we obtain a percentage of contact area of approximately 100 %, suggesting the graphene conforms to the Pd films.

We examine the quality of the graphene after evaporation or sputtering of Pd by Raman spectroscopy. To do so, we deposited a thin Pd film (5-10 nm) on graphene/$SiO_2$ by thermal evaporation or rf magnetron sputtering, and measured the Raman spectra of the Pd/graphene/$SiO_2$ samples using a home built Raman system with a 532 nm continuous laser. The positions of the G-peak of the samples deposited by thermal evaporation and rf magnetron sputtering are red-shifted to 1590 cm$^{-1}$ and 1595 cm$^{-1}$ respectively, see Figure 1d. From the magnitude of the shift, we estimate[26] a carrier concentration of $\approx$4-8×10$^{12}$ cm$^{-2}$ is induced in the graphene due to charge transfer from the Pd. We notice that for the samples prepared by thermal evaporation, there are no significant D peaks, see Figure 1d, suggesting that the graphene is undamaged after the thermal evaporation process, as we observed before.[22, 24] However, for the graphene samples prepared by rf magnetron sputtering, we observe a huge D peak, indicating the breakage of sp$^2$ bond, see Figure 1d. The ratio of intensity of D peak to 2D peak of the graphene sputtered at a rate of 0.3 Å/s and 1 Å/s is 2.8 and 5.9 respectively, suggesting the graphene sputtered at a rate of 1 Å/s is much severely damaged. We also observe broadening of the G peak for the sputtered samples, indicating the G peak has already merged with D' peak.[27, 28]

We measure the thermal conductance ($G$) of Pd/trG/Pd interfaces by time-domain thermoreflectance (TDTR). Details of our implementation[24, 29, 30], including our approach to eliminate the artifacts due to leaked pump beam[28], are discussed in our previous papers. We conducted the TDTR measurements using a 5x objective lens with $1/e^2$ radii of ≈10 μm. We used a total laser power of 80 - 200 mW to limit the steady-state temperature rise to ≈10 K. We extract the thermal conductance of the Pd/trG/Pd interfaces by comparing the ratio of in-phase and out-of-phase signals of TDTR measurements to calculations of a thermal model.[31] In the analysis, the thermal conductance of the Pd/trG/Pd interfaces is the only fitting parameters and all other parameters are obtained either from literature or stand-alone measurements, see reference 24 for details. We determine the thickness of the transducer film by picosecond acoustics.[32] However, the acoustic echoes observed in our experiments are weak due to the small piezo-optic coefficient of Pd. Therefore, we verify the thickness from AFM images of the thin films over sharp edges fabricated by photolithography, see Figure S3 in the supplementary. We then derive the thermal conductivity of the Pd films from the electrical resistivity (measured by four point probe) and the thickness of the Pd films that we obtained, using the Wiedemann-Franz law.

In this study, a major source of uncertainty is the thermal conductance of the bottom Pd/GaN interfaces. Hence, to improve the accuracy of our measurements, we prepared a Pd/GaN sample and measured the thermal conductance of the Pd/GaN interface by TDTR, for the same temperature range of $80 \leq T \leq 500$ K. We plot the thermal conductance of Pd/GaN interface and compare it with that of other metal/dielectrics interfaces in Figure 2b. We observe a weak temperature dependence, characteristic of interfaces in which phonons are the dominant heat carriers, see Figure 2b. Using the accurate thermal conductance of Pd/GaN, we successfully achieve an acceptable uncertainty of 9-23 % for our measurements of the thermal conductance of Pd/trG/Pd interfaces.

We summarize our measurements of the thermal conductance of Pd/trG/Pd interfaces over a temperature range of $80 \leq T \leq 500$ K in Figure 2a. At room temperature, we find that $G$=42 MW m$^{-2}$ K$^{-1}$ for Pd/trG/Pd interfaces prepared by thermal evaporation. This value of thermal conductance is comparable to prior measurements of the thermal conductance of metal contacts on transferred graphene (trG)[24, 33-35] and exfoliated graphene (exG)[22, 36] Similar to prior reported graphene interfaces[22, 24, 36, 37] and metal/dielectrics interfaces[21, 38-40] (including our data in Figure 2b), we observe a weak dependence of thermal conductance on temperature, which suggests that heat is primarily carried by phonons, and thus the electronic contribution is insignificant.

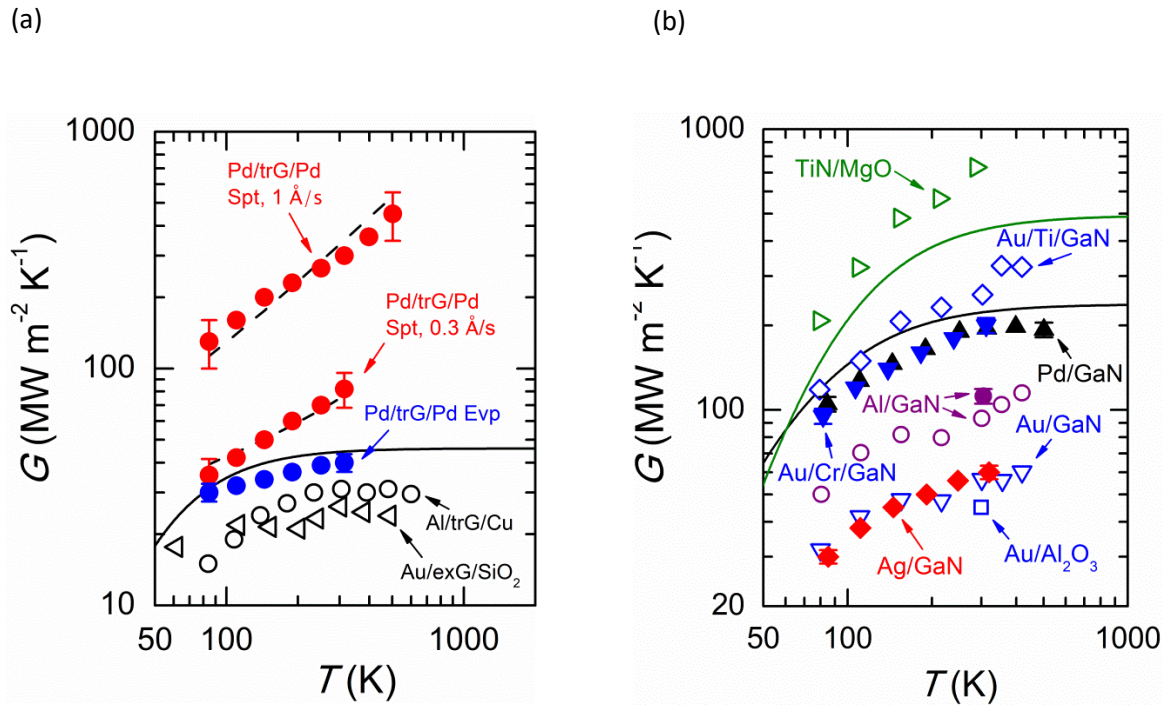

**Figure 2.** (a) Temperature dependence of the thermal conductance $G$ of interfaces of Pd/trG/Pd (solid circles, this work), compared to that of Al/trG/Cu (open circles, ref 25), Au/exG/SiO$_2$ (open left triangles, ref 23). The top metal layer of the Pd/trG/Pd interfaces was deposited either

by rf magnetron sputtering (red) or by thermal evaporation (blue). The solid line is calculations of the diffuse mismatch model (DMM) assuming that the thermal resistance of Pd/graphite and graphite/Pd interfaces adds in series. The dashed lines are fits of Eq. 3 to the thermal conductance measurements, using $\rho_c = 2.5 \times 10^{-10}$ Ω cm$^2$ and $2.2 \times 10^{-9}$ Ω cm$^2$, respectively. (b) Temperature dependence of the thermal conductance $G$ of Pd/GaN (solid black up triangles, this work), Au/Cr/GaN (solid blue down triangles, this work), Ag/GaN (solid red diamonds, this work), and Al/GaN (solid purple circle, this work), compared to the thermal conductance of interfaces of TiN/MgO (open olive right triangles, ref 21), Al/GaN (open purple circles, ref 38), Au/Ti/GaN (open blue diamonds, ref 38), Au/GaN (open blue down triangles, ref 38), and Au/Al$_2$O$_3$ (open blue square, ref 39). The dashed lines are the calculations of the DMM model for $G$ of Pd/GaN (black) and TiN/MgO (olive) interfaces, respectively.

The weak temperature dependence could be well explained by the diffuse mismatch model (DMM) for heat transport by phonons, see the comparison between our measurements and our DMM calculations in Figure 2. In our DMM calculations, we assume a truncated linear dispersion for the phonon dispersion[41] of Pd and employ the properties along the [100] direction. We assume that phonons are diffusely and elastically scattered at the interfaces and allow mode conversion at the interfaces. The transmission probability α can then be expressed as[22]

$$\alpha = \frac{I_2}{I_2+I_1} = \frac{I_2}{I_2+(\sum v_j^{-2})_1} \tag{1}$$

where $v_j$ is the speed of sound of phonons with mode $j$ and $I$ is the summation of $v_j$. We use subscript 1 to denote Pd side and subscript 2 to denote graphite side. Since graphite is anisotropic, we cannot determine the $I_{graphite}$ from the speeds of sound. Instead, we follow Koh et. al.[22] to use a fitted value of $I_{graphite} = 6.25 \times 10^{-8}$ s$^2$ m$^{-2}$. We then sum the thermal resistance of the top and bottom Pd/graphene interfaces in series and estimate $G_{Pd/graphene/Pd}$ from $G_{Pd/graphene/Pd}^{-1} = 2 G_{Pd/graphite}^{-1}$.

At first glance, our conclusion that electronic heat transport is negligible across Pd/graphene interfaces is surprising, since graphene is doped by Pd to a rather high level of ≈4-

$8\times10^{12}$ electrons/cm$^2$, and there is no native oxide on Pd. In this regard, past measurements on metal/metal interfaces in which two materials with a high concentration of electrons are in direct contact suggest that heat transport by electrons could be enormous (>1 GW m$^{-2}$ K$^{-1}$).[42] Moreover, Pd is chemisorbed on graphene, and thus the bonding strength of Pd/graphene is high.[43] To understand why electrons do not play a crucial role in heat transport across intrinsic metal/graphene interfaces, we consider the Wiedemann-Franz law for heat conduction by electrons across interfaces; $G_e = LT/\rho_c$, where $L=2.45\times10^{-8}$ Ω W K$^{-2}$ is the Lorenz number, $T$ is temperature and $\rho_c$ is contact resistivity. As suggested by the equation, the interfacial thermal conductance by electrons is mainly determined by the contact resistivity, not just the carrier concentrations in both sides of interfaces. For metal contacts on graphene, the contact resistivity is usually high even for the chemisorbed metals, due to the formation of Schottky-like barrier triggered by the mismatch in work functions of metals and the fermi level of charge carriers in graphene.[44]

Interestingly, we find that the thermal conductance of two Pd/trG/Pd interfaces prepared by rf magnetron sputtering, at a sputtering rate of 0.3 and 1.0 Å/s respectively, is significantly enhanced compared to that prepared by thermal evaporation. The thermal conductance of the sample prepared at a sputtering rate of 1.0 Å/s is 300 MW m$^{-2}$ K$^{-1}$ at room temperature, ≈7 times larger than that of the evaporated sample. We propose two possible explanations for the high thermal conductance of the sputtered samples. (1) First, due to unintentional treatment of Ar plasma during the sputtering process, Ar atoms could be adsorbed[45] on the surface of graphene. The adsorbed Ar atoms could strengthen the interface bonding between graphene and Pd, and thus enhance the heat conduction by phonons across the interfaces, similar to the enhancement of phononic heat transport across Al/graphene interfaces treated by oxygen plasma.[20] (2) Second, graphene could be damaged during the sputtering process because of bombardment of carbon atoms by Pd atoms. Similar damages were observed in graphene after deposition of oxides and nitrides by magnetron sputtering.[46, 47] Due to the partial removal of carbon atoms, electrons can transmit directly from the top to the bottom Pd films, and thus the thermal conductance could be enhanced due to an additional heat transfer channel by electrons.

To evaluate whether the observed enhancement of the thermal conductance is due to enhanced heat conduction by phonons (i.e., hypothesis #1) or electrons (i.e., hypothesis #2), we measure the thermal conductance of Pd/trG/Pd interfaces prepared by rf magnetron sputtering over a wide temperature range. We observe a strong dependence of thermal conductance on temperature, see Figure 2a. We then fit the temperature dependence measurements of the samples with

$$G = G_{ph} + LT/\rho_c \tag{2}$$

where $G_{ph}$ and $G_e = LT/\rho_c$ is the phononic (lattice) and electronic thermal conductance, respectively. Here, we postulate that the phononic component of the thermal conductance is not affected by damages in graphene and thus estimate $G_{ph}$ from the $G$ of the evaporated Pd/trG/Pd sample. We then assume that the contact resistivity $\rho_c$ is independent of temperature and fit the thermal conductance of the sputtered samples with equation (2). Our measurements agree well with the calculation using equation (2) over the entire temperature range, see Figure 2a, suggesting that the additional thermal conductance is due to an additional channel of heat conduction by electrons, instead of an increase in interfacial bonding strength.

Through the fits of equation (2), we estimate the contact resistivity $\rho_c$ to be $\approx 2.5 \times 10^{-10}$ $\Omega$ cm$^2$ and $\approx 2.2 \times 10^{-9}$ $\Omega$ cm$^2$ for the samples deposited at rate of 1.0 and 0.3 Å/s, respectively. The low contact resistivity could be due to either electron transport via direct Pd-Pd contacts or tunneling of electrons through atomic-scale gaps. If the size of the damages is sufficiently large, the Pd atoms in the top and bottom Pd layers are in direct contact, forming a metallic bridge for electrons to transmit. If the damages are atomic-scale, however, an atomic-scale gap of <0.5 nm can exist between Pd atoms on each side of graphene. In this case, the reduction in electrical contact resistivity is facilitated by tunneling of electrons through the atomic-scale pinholes. For a gap of 0.4 nm, the gap conductivity due to tunneling[48] is $10^3$ $\Omega^{-1}$ $cm^{-1}$ and thus the electrical contact resistivity is ≈4×$10^{-11}$ $\Omega$ $cm^2$, across the obliterated regions. Thus, provided that the area of the obliterated regions is ≈1 % of the total graphene area, the estimated contact resistivity of the damaged graphene through tunneling is consistent with the contact resistivity we observe in our samples. We are not able to pinpoint from our experiments whether the enhanced electron transmission is due to classical transport through metallic bridges or through tunneling across an atomic gap.

(a)
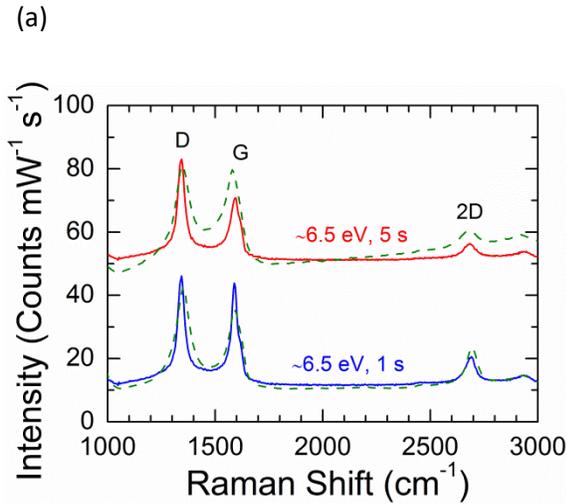

(b)
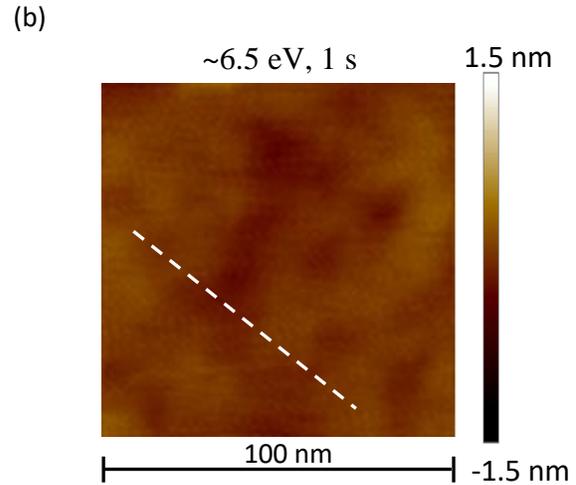

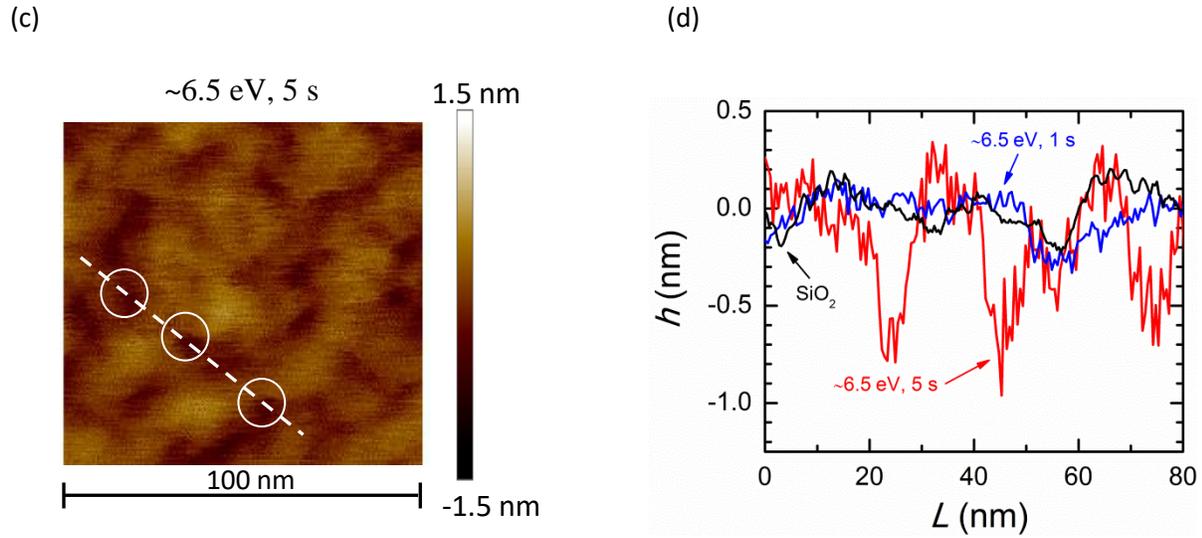

**Figure 3.** (a-c) Raman spectra and the corresponding AFM topographic images of transferred graphene on $SiO_2$ that was subsequently bombarded by argon ions of ~6.5 eV, for a duration of 1 s (b) and 5 s (c), respectively, as labeled. We control the dose of the ion bombardment by adjusting the duration of ion bombardment, such that the damage Ar ions induce in the graphene in (b) and (c) is comparable to the damage induced by rf magnetron sputtering of Pd with a deposition rate of 0.3 Å/s and 1Å/s respectively. The spectra of the sputtered graphene (green) are included in (a) for comparison. The AFM image of graphene in (b) is almost identical to that of pristine graphene, while the AFM image of graphene in (c) suggests that the graphene is damaged more severely compared to graphene in (b). The white circles are the possible regions with pinholes. (d) Comparison of relative height along the dashed line in (b) (blue) and height along the dashed line in (c) (red), with that of the $SiO_2$ substrate (black). The FWHM of the pinholes of the graphene bombarded by 6.5 eV ions for 5 s is ~7 nm.

To estimate the size of pinholes in the graphene of our sputtered Pd/trG/Pd samples, we transferred CVD graphene onto $SiO_2$ substrates and bombarded the trG/$SiO_2$ samples with biased argon ions. We kept the energy of the Ar ions at 6.5 eV and adjusted the dose of the ion bombardment via changing the duration of the bombardment (1 s and 5 s). We thus achieved two levels of damages in our trG/$SiO_2$ samples, with the Raman spectra similar to the Raman spectra of our sputtered samples, see Figure 3a. We employed tapping mode AFM with a sharp tip of 2 nm in nominal radius to measure the topographic profile of the ion-bombarded trG/$SiO_2$ samples.

For the trG/SiO$_2$ sample ion-bombarded for 5 s, we observe evident valleys which are unmistakably pinholes on graphene, see Figure 3. The full-width-half-maximum (FWHM) of the pinholes is estimated to be ≈7 nm. For the trG/SiO$_2$ sample ion-bombarded for 1 s, however, the AFM topographic profile is similar to that of a pristine sample. Our AFM images thus suggest that the pinholes generated by sputtering at a rate of 0.3 Å/s are atomic-scale with a FWHM much smaller than 2 nm.

In conclusion, we report the thermal conductance $G$ of Pd/trG/Pd prepared by both thermal evaporation and rf magnetron sputtering. We experimentally demonstrate that electrons do not play a significant role in heat transport across intrinsic metal/graphene interfaces, even when the graphene is highly doped by the metal. Interestingly, for the metal/graphene/metal interfaces, we observe significant heat transport by electrons when the graphene is damaged by sputtering. Moreover, for one of our samples, the electronic heat transport is still significant when the AFM topographic images suggest that the pinholes (i.e., damages) generated by the sputtering are atomic-scale (<2 nm). We propose that the enhanced electronic thermal transport could be due to either classical electron transport across metallic bridges or tunneling of electrons across the atomic-scale gaps created due to obliteration of carbon atoms by ion bombardment.

## Acknowledgements

This work is supported by NUS Young Investigator Award 2011, Singapore Ministry of Education Academic Research Fund Tier 2 under Award No. MOE2013-T2-2-147 and Singapore Ministry of Education Academic Research Fund Tier 1 FRC Project FY2016. Sample characterization was carried out in part in the Centre for Advanced 2D Materials.